\documentclass[letter]{aa} 
\usepackage{graphicx}
\usepackage{txfonts}

\begin{document}

   \title{Double mode RR Lyrae stars in Omega Centauri}

   \author{A. Olech
          \inst{1}
          \and
          P. Moskalik\inst{1}
          }

   \institute{Copernicus Astronomical Center, Polish Academy of Sciences,
              ul. Bartycka 18, 00-716 Warszawa, Poland\\
              \email{(olech,pam)@camk.edu.pl}
             }

   \date{Received December 1st, 2008; accepted ................., 2008}

\abstract
{}
{The aim of this work was to search for double mode pulsators among
 RR~Lyr variables of globular cluster $\omega$~Cen.}
{We conducted a systematic frequency analysis of CASE photometry of
 $\omega$~Cen RR~Lyr stars. We searched for periodicities using
 Fourier and ANOVA periodograms, combined with consecutive prewhitening
 technique.}
{We discovered six double mode pulsators, with the first overtone
 and a secondary mode of higher frequency simultaneously excited.
 These are the first double mode RR~Lyr stars identified in
 $\omega$~Cen. In variable V10 period ratio of the two modes is
 0.80, which corresponds to pulsations in the first and second
 radial overtones. In V19 and V105 we found unexpected period ratio
 of 0.61. Three other stars display period ratios of either $\sim\!
 0.80$ or $\sim\! 0.61$, depending on the choice of aliases.}
{While the period ratio of $\sim\! 0.80$ is easy to interpret in
 terms of two lowest radial overtones, the value of $\sim\! 0.61$
 cannot be explained by any two radial modes. Thus, V19 and V105 are
 the first members of a new class of double mode RR~Lyr pulsators.}

\keywords{stars: oscillations -- stars: horizontal branch --
          globular clusters: individual: $\omega$~Centauri}

\maketitle

\section{Introduction}

The RR~Lyr type pulsating stars are one of the most interesting and
most important objects in astrophysics, currently representing one of
the basic rungs in the "distance ladder". Already from the moment of
their discovery it was realized that their light curves were not
uniform and could be divided into three classes: a, b and c (Bailey
1902). Later these classes were simplified to two types: RR$ab$
(radial pulsations in the fundamental mode) and RR$c$ (radial
pulsations in the first overtone).

For the next 75 years it appeared that these were the only forms
of pulsations possible in this type of stars. However, in the late
1970s and early 1980s it was found that RR~Lyr stars can pulsate
with both fundamental mode and the first overtone simultaneously
excited (Jerzykiewicz \& Wenzel 1977; Cox et al. 1983). Later it
was discovered that in addition to radial pulsations, RR~Lyr stars
can also display non-radial modes of similar periods (Olech et al.
1999). Possibility of pulsations in the second radial overtone was
discussed in the literature as well. For example, Alcock et al.
(1996) found three maxima in the period distribution of RR~Lyr
stars in the LMC and proposed that the shortest period maximum
corresponds to the second overtone. Cluster RR~Lyr stars with
nearly sinusoidal, low amplitude light curves and very short
periods were interpreted as second overtone pulsators, too ({\em
e.g.} Walker \& Nemec 1996, Kaluzny et al. 2000). However,
presented evidence was weak, and all these stars could be
pulsating in the first overtone after all. Theoretical models
actually predict existence of such short period, low amplitude
RR$c$ stars ({\em e.g.} Bono et al. 1997).

While in RR~Lyr stars there is no conclusive evidence for
excitation of the second overtone, such a mode is observed in many
Cepheids. In the the latter case, unambiguous evidence for single
mode second overtone pulsations have been found only recently
(Soszy\'nski et al. 2008), but double mode Cepheids with the first
and second overtones simultaneously excited have been known for
years (Mantegazza 1983; Alcock et al. 1995).

\begin{table*}[!t]
\caption[]{Pulsation frequencies of V10, V350, V81, V87, V105 and V19.}
\centering
\begin{tabular}{|l|c|c|c|c|c|c|}
\hline
\hline
Star                 & $P_1$\thinspace [day]
                                      & $P_2$\thinspace [day]
                                                      & $P_2/P_1$ & $A_1$\thinspace [mag]
                                                                                & $A_2$\thinspace [mag]
                                                                                              & $\sigma$\thinspace [mag] \\
\hline
V10 ~~~~1st\,\, sol. & ~0.3749759(5)~ & ~0.299176(9)~ & ~0.79785~ & ~0.1831(4)~ & ~0.0067(4)~ & ~0.0071~ \\
V10 ~~~~2nd     sol. & ~0.3749758(5)~ & ~0.230126(6)~ & ~0.61371~ & ~0.1825(5)~ & ~0.0064(5)~ & ~0.0074~ \\
\hline
V350 ~~1st\,\, sol.  & ~0.3791074(3)~ & ~0.230632(5)~ & ~0.60836~ & ~0.2126(6)~ & ~0.0062(6)~ & ~0.011~ \\
V350 ~~2nd     sol.  & ~0.3791077(4)~ & ~0.303810(9)~ & ~0.80138~ & ~0.2117(6)~ & ~0.0058(6)~ & ~0.011~ \\
V350 ~~3rd\,   sol.  & ~0.3749758(5)~ & ~0.230126(6)~ & ~0.61371~ & ~0.1825(5)~ & ~0.0064(5)~ & ~0.011~ \\
\hline
V81 ~~~~1st\,\, sol. & ~0.3893907(3)~ & ~0.238990(4)~ & ~0.61375~ & ~0.2214(5)~ & ~0.0072(5)~ & ~0.010~ \\
V81 ~~~~2nd     sol. & ~0.3893907(3)~ & ~0.314313(7)~ & ~0.80719~ & ~0.2217(5)~ & ~0.0065(6)~ & ~0.010~ \\
\hline
V87 ~~~~1st\,\, sol. & ~0.3964889(3)~ & ~0.246584(4)~ & ~0.62192~ & ~0.2314(6)~ & ~0.0063(5)~ & ~0.010~ \\
V87 ~~~~2nd     sol. & ~0.3964884(3)~ & ~0.320542(8)~ & ~0.80845~ & ~0.2322(6)~ & ~0.0055(6)~ & ~0.010~ \\
\hline
V105                 & ~0.3353278(2)~ & ~0.205811(2)~ & ~0.61376~ & ~0.2465(6)~ & ~0.0125(6)~ & ~0.010~ \\
\hline
V19                  & ~0.2995510(2)~ & ~0.183302(2)~ & ~0.61192~ & ~0.2279(5)~ & ~0.0071(5)~ & ~0.010~ \\
\hline
\hline
\end{tabular}
\end{table*}

\begin{figure*}
 \vspace{4cm}
 \caption{Double mode RR~Lyr variable V10. (a) light curve phased with the
          primary period $P_1$, (b) power spectrum of original light curve,
          (c) power spectrum of prewhitened light curve.}
\includegraphics{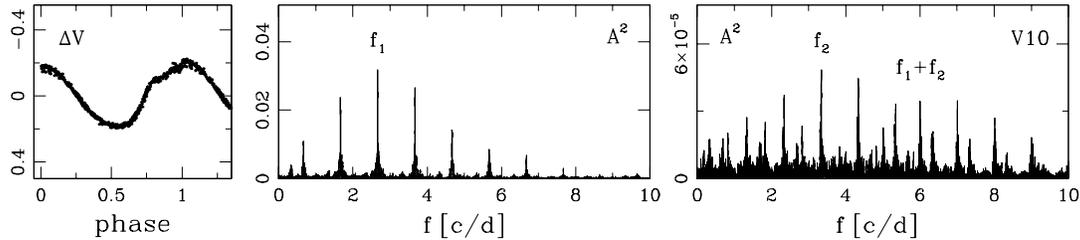}
\end{figure*}

\begin{figure*}
 \vspace{11.6cm}
 \caption{Same as Fig.\thinspace 1, but for double mode RR~Lyr variables V350,
          V81, V87, V105 and V19.}
\includegraphics{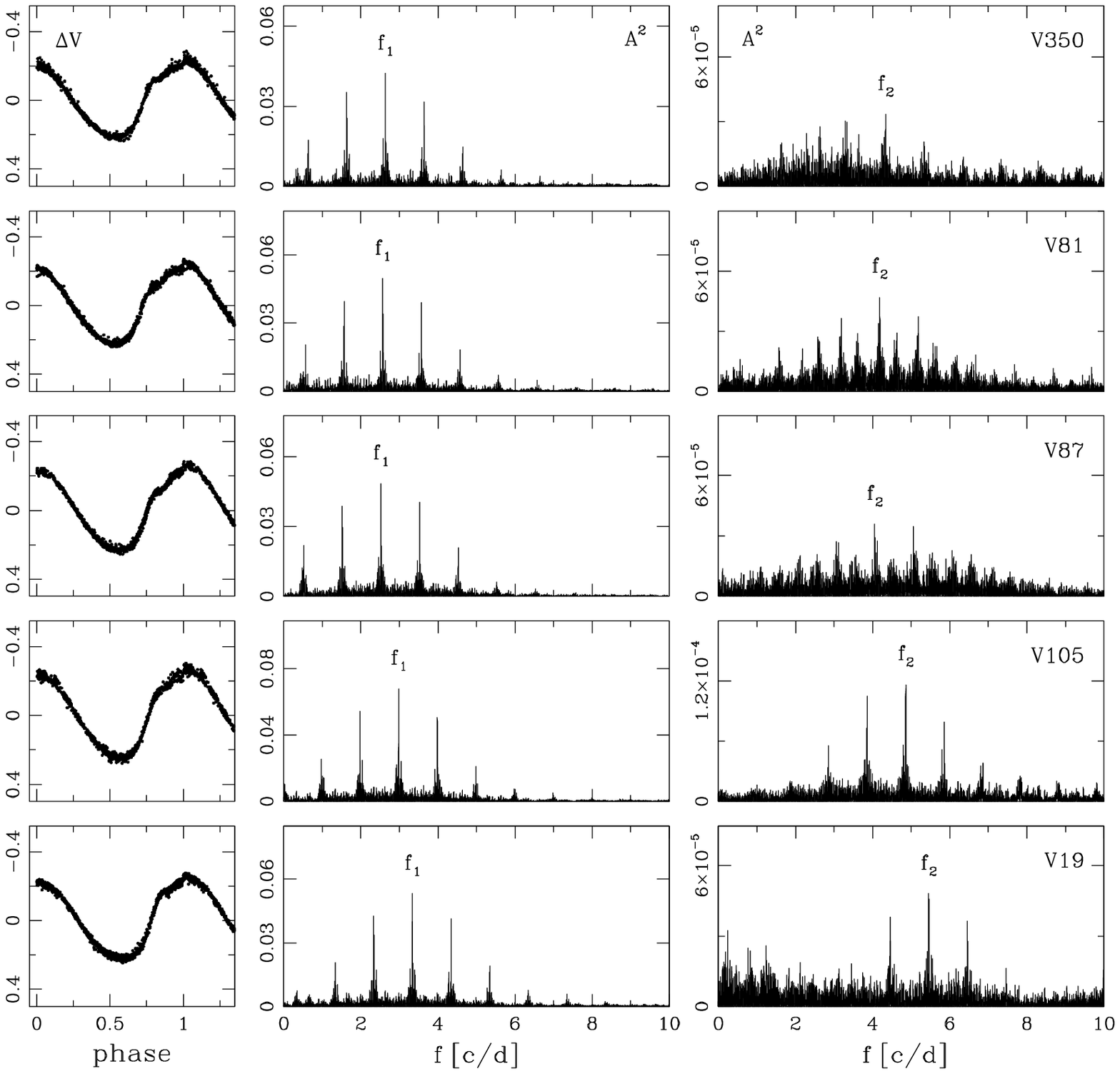}
\end{figure*}

\section{Data and frequency analysis}

Omega Centauri ($\omega$~Cen) is the largest globular cluster of the
Galaxy. It consists of about one million stars, of which almost 500
are known to be variable (Kaluzny et al. 2004, Weldrake et al.
2007). Very accurate $B$ and $V$ CCD photometry of $\omega$~Cen
variables has recently been published by the Cluster AgeS Experiment
or CASE (Kaluzny et al. 2004). The paper provides high quality
photometry of 151 RR~Lyr pulsators. The data have been collected
between February 6/7, 1999 and August 9/10, 2000 and contained from
594 to 761 points per star.

We conducted a systematic frequency analysis of all RR~Lyr stars
observed by the CASE project. This was done with two different methods.
In both cases we used the standard consecutive prewhitening technique
(Moskalik et al. 2004). In the first approach, the data were first
fitted with the single frequency Fourier sum of the form:

\vskip 5pt

\begin{equation}
m(t) = A_{0} + \sum_{k} A_{k} \sin (2\pi k f_{0} t + \phi_{k})
\end{equation}

\noindent where the primary pulsation frequency $f_{0} = 1/P_{0}$
was also optimized in the fitting process. The fit Eq.(1) was then
subtracted from the data and the residuals were searched for
secondary periodicities. This was done with the Fourier power
spectrum computed over the range of $0-10$ c/d. If any additional
signal was detected, a new Fourier fit with all frequencies
identified so far and with their linear combinations was performed.
All frequencies were optimized anew. The new fit was subtracted from
the original time-series data and the residuals were searched for
additional periodicities again. The process was repeated until no
new frequencies appeared. At this stage we performed data clipping,
by rejecting all measurements deviating from the fit by more than
5$\sigma$, where $\sigma$ was dispersion of the residuals. After
removing deviating datapoints, the entire frequency analysis was
repeated.

Our second approach differed in treatment of residuals. We started
with the ANOVA periodogram (Schwarzenberg-Czerny 1996) of the raw
time-series data and searched it for the peak frequency. The data
were then prewhitened with the frequency just identified. Depending
on the amplitudes, we removed up to 15 harmonics of the frequency
from the {\em prewhitened data} of the previous stage. Next, we
recomputed the ANOVA periodogram for the residuals and the whole
procedure was repeated until no feature exceeding ANOVA value of 15
appeared.

Both methods yielded the same results within statistical errors. The
only discrepancies appeared when a true frequency and its alias had
similar amplitudes and the first method preferred one peak while the
second method preferred the other. In such a cases we chose the
frequency, which yielded a fit with lower $\sigma$.

\section{Results}

\subsection{Periods and period ratios}

In the course of our analysis we identified many RR~Lyr variables
with secondary periodicities close to the primary (radial) pulsation
frequency. These objects, commonly referred to as Blazhko RR~Lyr
stars, are discussed elsewhere (Moskalik \& Olech 2008; 2009). In
six of the RR$c$ variables we detected multiperiodicity of a
different kind -- a secondary mode appeared at a frequency much
higher than the primary one. These stars are listed in Table~1.

The most interesting case is variable V10. Its primary period of
$P_1=0.3749759$\thinspace day and its light curve are typical for
RR$c$ pulsator. Prewhitening led to discovery of the second
frequency, corresponding to the period of $P_2=0.299176$\thinspace
day. The resulting period ratio of $P_2/P_1=0.79785$ is
characteristic for simultaneous pulsation in the first and second
overtone (FO/SO double mode variable).

\begin{figure}
 \vspace{7.5cm}
 \caption{Color-magnitude diagram for horizontal branch of $\omega$~Cen. Filled
          and open circles correspond to RR$ab$ and RR$c$ star, respectively.
          Nonvariable stars are displayed with dots. Triangles mark double mode
          RR~Lyr variables of Table~1}
\includegraphics{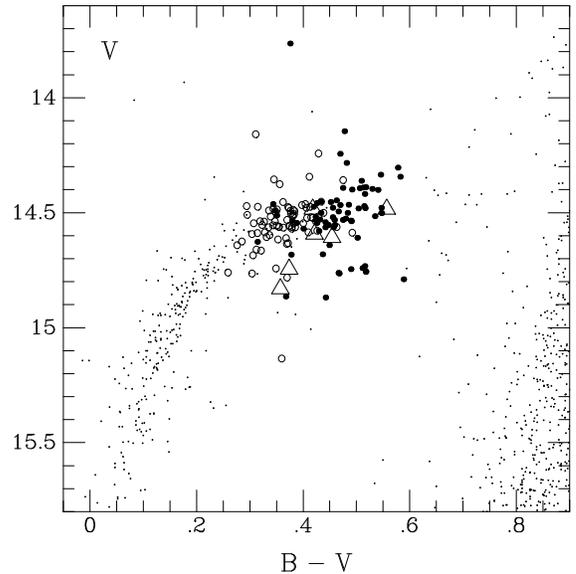}
\end{figure}

The light curve of V10, its Fourier power spectrum and power
spectrum of the prewitened light curve are shown in Fig.\thinspace 1.
Both $f_2$ and the combination peak $f_1+f_2$ ale clearly visible.
We note, that the spectral window is not very good and it is
possible that the true secondary frequency is at the 1-day alias of
the highest peak. If this is the case, then $P_2 =
0.230126$\thinspace day, giving period ratio of $P_2/P_1=0.61371$.
This solution is listed in the second line of Table~1. It yields
higher dispersion of the least-square fit, but it cannot be
definitely excluded. Nevertheless, the most likely period ratio in
V10 is $\sim\! 0.80$, which makes this star a strong candidate to be
the first FO/SO double mode pulsator among RR~Lyr variables.

\begin{figure}
 \vspace{7.5cm}
 \caption{Period-amplitude diagram for RR~Lyr stars of $\omega$~Cen. Same
          symbols as in Fig.\thinspace 3}
\includegraphics{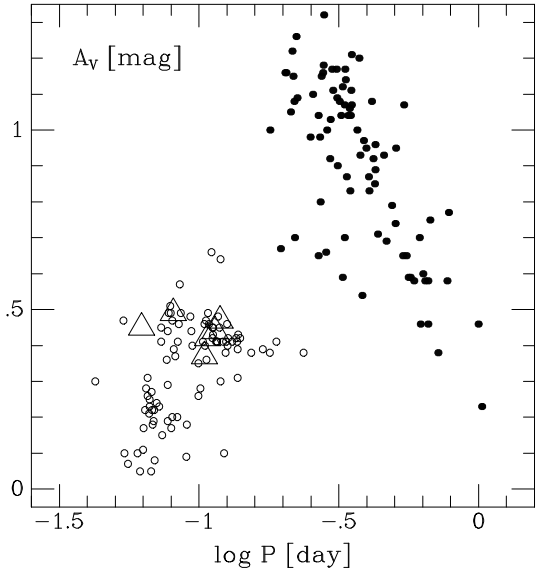}
\end{figure}

We found five more double mode pulsators with high frequency
secondary modes. Their properties are summarized also in Table~1. In
Fig.\thinspace 2 we display light curves, original power spectra and
power spectra of prewitened light curves of these stars. Judging
from shapes of their light curves, there is no doubt that in all
five stars the primary frequency corresponds to the first overtone.
There is also no doubt that secondary frequencies are real. The
first three star, V350, V81 and V87, are similar to V10: they
display the period ratios of either $\sim\! 0.80$ or $\sim\! 0.61$,
depending on the choice of an alias. This time, selecting the true
alias is not possible, however, because both choices lead to
least-square fits of the same quality. We were initially tempted to
reject the period ratio of $\sim\! 0.61$ as unphysical. As it turned
out, this would be unjustified.

The most surprising result was found for the other two double mode
pulsators, V105 and V19. For these two variables we derived a period
ratio of $P_2/P_1\sim\! 0.61$ and it was {\em the only acceptable
possibility}. This puzzling value does not correspond to any two
radial modes. Thus, variables V105 and V19 constitute a new type of
double mode RR~Lyr pulsators.

\subsection{Color-magnitude diagram}

\begin{figure*}
 \vspace{4.4cm}
 \caption{The amplitudes ratios $R_{j1}$ and phase
          differences $\phi_{j1}$ {\em vs.} first overtone period $P$ for all
          RR$c$ variables of our sample. Triangles mark double mode pulsators
          of Table~1.}
\includegraphics{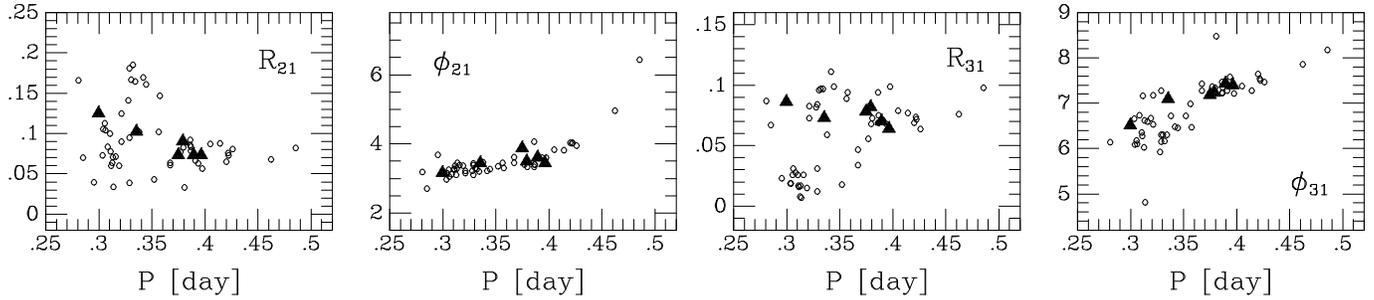}
\end{figure*}

Fig.\thinspace 3 presents the $V-(B\!-\!V)$ color-magnitude
diagram for horizontal branch of $\omega$~Cen. It is worth noting
that RR$c$ and RR$ab$ regions overlap significantly. There is a
clear scatter in mean $V$ magnitude of the horizontal branch
stars, which is most likely caused by differences in their
chemical composition. It is known that stars of $\omega$~Cen can
differ in metallicity by factor of up to 100 (Sollima et al.
2005). Six double mode variables discussed in the previous section
are displayed with filled triangles. Because they are strongly
dominated by the first overtone, we will compare them with the
RR$c$ stars. Five of the double mode variables are located on the
cooler side of the RR$c$ region, but without any significant trend
or clumping. The clear exception is V350, which seems to be as red
as the redest fundamental mode pulsators. This variable is a new
detection made by Kaluzny et al. (2004) with the image subtraction
technique. It is located in a very crowded region, where
significant blending can affect both mean magnitude of the star
and its color.

\subsection{Period-amplitude diagram}

Typical period-amplitude diagram for RR~Lyr stars of a globular
cluster shows roughly linear trend for RR$ab$ stars, with smaller
amplitudes at longer periods, and a clump of RR$c$ stars with no
clear structure. Some investigators try to divide this clump by
amplitude into two subgroups, containing second and first overtone
pulsators, respectively ({\em e.g.} Clement \& Rowe 2000). This is
not well justified, as the shortest period first overtone pulsators
can have very small amplitudes as well (Bono et al. 1997).

The period-amplitude diagram for RR~Lyr stars of $\omega$~Cen is
shown in Fig.\thinspace 4. The double mode pulsators are plotted
with filled triangles. All of them are located firmly among RR$c$
variables. We note, however, that they all have high pulsation
amplitudes, one of the highest among overtone pulsators.

\subsection{Fourier coefficients}

Fourier coefficients (amplitudes $A_j$, amplitude ratios $R_{ij}$ and
phase differences $\phi_{ij}$) could be used to characterize the shape
of the pulsation light curves and to discriminating between different
modes of pulsation. We decided to compare Fourier coefficients of our
six double mode variables with those of the RR$c$ pulsators. The results
are shown in Fig.\thinspace 5. In all cases, the position of the double
mode stars is typical for RR$c$ variables, without any trend or
clumping. It confirms conclusions which could be drawn form the visual
inspection of the light curves of the variables. They look quite typical
for RR$c$ stars, without any significant pecularity, indicating that
influence of the seconadry mode is weak and this mode does not  affect
the global shape of the light curve. The only surprising thing is large
pulsation amplitude which was mentioned in the previous paragraph.

\section{Conclusions}

Systematic frequency analysis of RR~Lyr stars of globular cluster
$\omega$~Cen resulted in discovery of six RR$c$ variables, which
in addition to the dominant first radial overtone, display also a
weak secondary mode of higher frequency. These are the first
double mode RR~Lyr stars identified in this cluster. In variable
V10 the most probable period ratio of the two modes is 0.80. This
value points towards pulsations in the first and second radial
overtones. In three other stars, the period ratios are either
$\sim\! 0.80$ or $\sim\! 0.61$, depending on the choice of
aliases. Finally, in the last two stars (V19 and V105) an
unambiguous period ratio of 0.61 was found. Such a period ratio
cannot be explained by two radial modes, which implies that the
secondary mode must be nonradial ({\em cf.} Moskalik \&
Ko{\l}aczkowski 2009 for discussion of this point). Thus, V19 and
V105 belong to a new class of double mode RR~Lyr pulsators. We
recall, that a similar period ratio of $\sim\! 0.61$ was also
discovered in the LMC first overtone Cepheids (Moskalik \&
Ko{\l}aczkowski 2008; Soszy\'nski et al. 2008) and in the field
double mode RR~Lyr star AQ~Leo (Gruberbauer et al. 2007).

Discovery of RR~Lyr stars pulsating in the first and second
overtones has been claimed previously in the LMC (Alcock et~al.
2000). However, Soszy\'nski et~al. (2003) have shown, that all
these objects are about 1\thinspace mag brighter than typical
RR~Lyr stars in the LMC, and suggested that they might be
short-period Cepheids instead. In contrast, all six double mode
variables discovered in $\omega$~Cen belong to the RR~Lyr
population of the cluster without any doubt. Thus, V10 is the
first solid candidate for the FO/SO double mode RR~Lyr pulsator.

\begin{acknowledgements}
This work was supported by MNiI grant number 1 P03D 011 30 to P.M.
\end{acknowledgements}

\end{document}